\documentclass[12pt]{article}
\usepackage{epsfig}
\usepackage{float}
\usepackage{amssymb,amsmath}
\newcommand{\be}{\begin{equation}}
\newcommand{\ee}{\end{equation}}
\newcommand{\bea}{\begin{eqnarray}}
\newcommand{\eea}{\end{eqnarray}}
\begin{document}
\begin{center}
 {\bf Comment on ''Hypersharp Resonant Capture of Neutrinos as a Laboratory Probe of the Planck Length''}
\end{center}
\begin{center}
W. Potzel and F. E. Wagner
\end{center}
\begin{center}
{\em Physik-Department E15, Technische Universit\"at M\"unchen 
\\D-85748 Garching, Germany}
\end{center}

\begin{abstract}

In \cite{Raghavan}, R.S. Raghavan claims that due to motio\-nal averaging by lattice vibrations, 18.6 keV $\bar{\nu_{e}}$ emitted/captured without recoil from 2-body decay in the $^{3}$H/$^{3}$He system embedded in Nb metal will be observable with natural width $\Gamma_{nat}$. In this comment we argue that
1) stochastic relaxation processes and 2) inhomogeneities in the metal matrices will prevent the generation of antineutrinos with $\Gamma_{nat}$ in the $^{3}$H/$^{3}$He system,
3) the different lattice-deformation energies of $^{3}$H and $^{3}$He in the Nb matrix will drastically decrease the fraction of phononless emission/detection of antineutrinos,
4) the age itself of the $^{3}$H source does not affect the linewidth.
\end{abstract}

\section{Stochastic Relaxation Processes}
Harmonic lattice vibrational motions do not cause a broadening of the M\"oss\-bauer line. Raghavan demonstrates this by using a continuous frequency modulation (FM) model. However, for $^{3}$H/$^{3}$He embedded in a metal (like Nb) there exist stochastic magnetic relaxation effects which are \textit{not} connected to lattice vibrations.
Spin-spin interactions between nuclear spins of $^{3}$H and $^{3}$He and with the spins of the Nb nuclei of the Nb metallic lattice lead to fluctuating magnetic fields. A simple relaxation model consists of the groundstate and two excited hyperfine-split states (separation $\hbar\omega_{0}$).  Stochastic processes are characterized by sudden and irregular transitions between these hyperfine-split states with an \textit{average} rate $\omega$. Stochastic transitions can not be described by a continuous FM model. It has been known for a long time (see., e.g., \cite{Wickman}) that stochastic relaxation processes do increase the linewidth if $\omega$ is comparable to $\omega_{0}$.
For typical magnetic hyperfine splittings due to nuclear spin-spin interactions in metallic lattices, $\omega_{0}\approx10^{5}$ s$^{-1}$, and typical relaxation times for $^{3}$H and $^{3}$He in a metallic Pd lattice are $T\approx2$ ms, and for NbH, $T\approx79$ $\mu$s \cite{Stoll}. Thus, for the system $^{3}$H/$^{3}$He in Nb metal, $\omega_{0}\approx \omega=2\pi/T$  and experimental linewidths of $\Gamma_{exp}=\hbar\omega\approx50\times10^{-12}$ eV $\approx4\times10^{13} \Gamma_{nat}$ have to be expected. As a consequence, the resonance cross-section will be reduced by about 13 orders of magnitude!

\section{Inhomogeneities in the metal matrices}

Inhomogeneities are caused by lattice defects, variations in the lattice constant, impurities, etc. Such inhomogeneities and, in particular, the highly different concentrations and random distributions of $^{3}$H and $^{3}$He in source and target will cause variations of the binding energies $E_B$ of $^{3}$H and $^{3}$He atoms in Nb metal. With the conventional M\"ossbauer effect (ME), different $E_B$ values due to inhomogeneities influence the photon energy only through the \textit{difference} of the mean-square nuclear charge radius between the groundstate and the excited state in the \textit{same type} of nucleus in source and target. This causes a shift of the photon energy typical for hyperfine interactions ($\approx10^{-8}$ eV). In the best single crystals, inhomogeneous broadening is $10^{-13}$ to $10^{-12}$ eV \cite {WPotzel}. With $\bar{\nu_{e}}$ emission and capture, however, \textit{nuclear transformations} occur. In the Nb lattice, $E_B$ per atom for $^{3}$H and $^{3}$He is in the eV range \cite{Puska}, many orders of magnitude larger than hyperfine interaction energies. Since, in the nuclear transformations, the $\bar{\nu_{e}}$ energy is \textit{directly} affected by variations of $E_B$, one has to expect a variation of the $\bar{\nu_{e}}$ energy by several orders of magnitude larger than $10^{-12}$ eV, i.e., larger than $10^{12}\Gamma_{nat}$.

\section{Decrease of the fraction of phononless emission and detection}

The different lattice-deformation energies $E_L$ for $^{3}$H and $^{3}$He in the Nb lattice \cite{Puska} have the consequence that $E_L$ changes by $\approx0.45$ eV at the lattice site where the nuclear transformation occurs. If the lattice rearrangement is accompanied by phonon generation, the resonance condition for the ME to occur will be destroyed. An estimate gives a reduction factor of $1\cdot10^{-6}$ for phononless emission and capture of the $\bar{\nu_{e}}$ \cite {WPotzel}. The argument given in \cite{Raghavan} that such lattice excitations are harmless since they occur only with the speed of sound does not hold for the conventional (photon) ME and is also not valid in the case of $\bar{\nu_{e}}$ interactions.

\section{Different physical ages of the source}

If the Fourier transform of an exponentially decaying oscillation (decay constant $\Gamma$) starting at $t=0$ is $F(\Omega)$, then the Fourier transform of such a decay, but taking into account only the time from $t=t_{0}>0$ to $t=\infty$, is given by $F(\Omega)\cdot e^{-(\Gamma/2)t_0}$. It has the same width as $F(\Omega)$, only the amplitude is reduced. Thus, an exponential decay has no memory of when it started. The decay of $^{3}$H is exponential.
In a $\bar{\nu_{e}}$ M\"ossbauer experiment, the clock is started together with the measurement, i.e., at the moment when source ($^{3}$H) and detector ($^{3}$He) are arranged in their fixed positions. The measuring time, not the age itself of the $^{3}$H source, is important for the linewidth.

\section{Acknowledgements}
This work was supported by funds of the Munich Cluster of Excellence (Origin and Structure of the Universe), the DFG (Transregio 27: Neutrinos and Beyond), and the Maier-Leibnitz-Laboratorium in  Garching.


\begin{thebibliography}{99}

\bibitem{Raghavan} R. S. Raghavan, Phys. Rev. Lett. \textbf{102}, 091804 (2009), preprint arXiv: hep-ph/0903.0787.


\bibitem{Wickman} H. H. Wickman and G. K. Wertheim, in \textit{Chemical Applications of M\"ossbauer Spectroscopy}, V. I. Goldanskii and R. H. Herber, eds. (New York: Academic Press, 1968), pp. 548, in particular Fig. 11.10.


\bibitem{Stoll} M. E. Stoll and T. J. Majors, Phys. Rev. \textbf{B24}, 2859 (1981) and references therein.



\bibitem{WPotzel} W. Potzel, J. Phys. Conference Series \textbf{136}, 022010 (2008), and references therein, preprint arXiv: 0810.2170.

\bibitem{Puska} M. J. Puska and R. M. Nieminen, Phys. Rev. B \textbf{29}, 5382 (1984).





\end{thebibliography}
\end{document}